\documentstyle[12pt]{article}

\begin{document}
\pagestyle{empty}
$\ $

\rightline{CWRU-P3-1995}

\rightline{VAND-TH-95-1}

\vskip 1.5 truecm

\centerline{\bf
Topological Incarnations of Electroweak Defects
}

\

\centerline {Thomas W. Kephart}
\smallskip
\centerline
{\it
Department of Physics and Astronomy, Vanderbilt University}
\centerline
{\it
Nashville, TN 37235.
}

\

\centerline
{Tanmay Vachaspati}
\smallskip
\centerline
{\it
Physics Department, Case Western Reserve University}
\centerline
{\it
Cleveland, OH 44106.
}

\vskip 1. truecm

\begin{abstract}
We propose a criterion to classify hybrid defects occuring in
field theoretic models such as the standard electroweak model.
This criterion leads us to consider the minimal extension of
the electroweak model in which electroweak magnetic monopoles
and $Z$-strings are topological. We briefly discuss the cosmology
of such defects.

\end{abstract}
\clearpage
\pagestyle{plain}

The standard electroweak model is based on an $SU(2)_L\times U(1)_Y$
symmetry group with the order parameter $\Phi$ being an $SU(2)_L$
doublet. When $\Phi$ acquires a vacuum expectation value (vev), the
symmetry breaks down to the electromagnetic $U(1)_Q$. Given this
symmetry breaking scheme, one finds that the vacuum manifold is
a three sphere which has trivial first and second homotopy groups.
This means that there are no topological monopoles and strings in
the electroweak model. However,these topological considerations are
blind to any {\it confined defects} \cite{trebin} that may be present
in the model.

It is known that the standard model of the electroweak interactions
contains confined magnetic monopoles \cite{yn}. These
magnetic monopoles are confined by Nielsen-Olesen strings \cite{hnpo}
which carry magnetic flux of the $Z$ boson \cite{yn, nm, tv}. The
existence of the $Z$-strings can be viewed as being due to the
embedding of topological defects within the standard model
\cite{nm, tvmb, mbtvmb}. But this viewpoint does not apply to
the monopole and a somewhat different reasoning
must be used to understand the existence of magnetic monopoles
in the standard model.

How might we detect confined defects from the topology of the
model? A topological scheme which is sensitive to confined defects
follows once we realize that when an order parameter acquires a vev,
composite operators (suitably normal ordered) that can be built from
the order parameter can also acquire vevs. The vev of these composite
operators yields different symmetry breaking patterns with vacuum
manifolds which have different topological properties. If the vacuum
manifold resulting from the vev of a composite operator has non-trivial
first or second homotopy groups, we can get confined strings or
confined monopoles.

This scheme for classifying confined defects can be illustrated in
the case of the standard model. Here, when $\Phi$
acquires a vev, the composite operator
\begin{equation}
\vec \chi = {\sqrt{2} \over \eta} :\Phi^{\dag} {\vec \tau} \Phi :
\label{chi}
\end{equation}
also acquires a vev. (We have used standard notation such as given in
\cite{chengli} and normal ordered with respect to the $\Phi$ ground
state vacuum). The field $\vec \chi$ transforms in the adjoint
representation of $SU(2)_L$ and is a singlet under $U(1)_Y$. So
the vev of $\vec \chi$ breaks $SU(2)_L$ to $U(1)'$ and the vacuum
manifold is $S^2$ with a non-trivial second homotopy group. Therefore
there are magnetic monopoles present in the model. In other words,
Nambu's electroweak magnetic monopole is precisely a hedgehog in
the $\vec \chi$ field \cite{gtap}.

The reason the monopoles are confined is that the relation (\ref{chi})
for a hedgehog configuration of $\vec \chi$ cannot be globally inverted
to give a non-vanishing field configuration for $\Phi$. Therefore
there is a line emanating from the monopole along which $\Phi$
vanishes. This is the location of the $Z-$string attached to the
monopole. The $Z-$string is {\it not} a Dirac string that returns the
electromagnetic flux; it is a string that confines the magnetic
monopole without having anything to do with the electromagnetic
flux.

Even higher dimensional
representation operators can be constructed from $\Phi$
and these will yield other confined defects. For example, the 5
dimensional representation of $SU(2)_L$ constructed using
$( \Phi^{\dag} {\vec \tau} \Phi )^2$ will
break $SU(2)_L \times U(1)_Y$ to $U(1)\times Z_2 \times U(1)_Y$
yielding an even richer array of confined topological defects such
as Alice strings bounded by walls. Here we will not discuss this very
interesting issue further but restrict our attention to electroweak
magnetic monopoles and $Z-$strings.

Having realised that the monopoles are hedgehogs in the composite
real scalar field $\vec \chi$, we can promote the monopoles to being
truly topological objects by promoting $\vec \chi$ to be a fundamental
field. So the Lagrangian we will consider is:
\begin{equation}
L = T_{ew} + |(\partial_\mu +ig \epsilon^a W_\mu ^a ) {\vec \chi}|^2
      - V(\Phi , {\vec \chi}) + L_f\
\label{lag}
\end{equation}
where, $T_{ew}$ is the gradient part of the bosonic sector of the
electroweak Lagrangian, $L_f$ is the fermionic part of the
Lagrangian, $\epsilon^a_{ij} = \epsilon_{aij}$ ($a,i,j = 1,2,3$)
and,
\begin{equation}
V(\Phi , {\vec \chi}) =
- \mu_2^2 \Phi^{\dag}\Phi - \mu_3^2 {\vec \chi}^2
+ \lambda_2 (\Phi^{\dag}\Phi)^2
+\lambda_3 {\vec \chi}^4 + a {\vec \chi}^2 \Phi^{\dag}\Phi
+ b {\vec \chi} \cdot \Phi^{\dag} {\vec \tau} \Phi \; .
\label{pot}
\end{equation}
If we impose an additional $Z_2$ symmetry on the Lagrangian under
$$
\Phi \rightarrow \Phi \ , \ \ \
{\vec \chi} \rightarrow -{\vec \chi} \ .
$$
the symmetry is $SU(2)_L\times U(1)_Y \times Z_2$ and
we must set $b=0$. In what follows, we shall only consider this
case and henceforth ignore the last (cubic) term in the
potential \cite{comment0}.
In this case, the vev of $\Phi$ breaks the symmetry
down to $U(1)_Q \times Z_2$.
Note that the leptons and quarks do not couple to $\vec \chi$ since
we have imposed the $Z_2$ symmetry on $L$.
This is because there is no Lorentz invariant dimension four operator
that can be constructed using the left-handed fermion $SU(2)$ doublets
and the ${\vec \chi}$ field which is a triplet and still be invariant
under the $Z_2$ transformation. Therefore $L_f$ is identical
to the fermionic Lagrangian of the standard electroweak model.

At non-zero temperatures, the coefficients in the potential will get
thermal corrections. The lowest order corrections give:
\begin{equation}
\triangle V = T^2 \left\{ \beta_2 \Phi^{\dag}\Phi
+ \beta_3 {\vec \chi} \cdot {\vec \chi} \right\}
 \label{deltaV}
\end{equation}
where
\begin{equation}
\beta_2 = {1 \over {48}} \left[ 6(4 \lambda_2 + a)+
3(3g^2+g^{\prime2})  + 4 \sum_{f} N_C G^{*}_{2f}G_{2f} \right]
 \label{beta2}
\end{equation}
and
\begin{equation}
\beta_3 = {1 \over 48} \left[ 4 (5\lambda_3 + 2 a)
+ 3g^2 + 4 \sum_{f} N_C G^{*}_{3f} G_{3f} \right] \quad .
\label{beta3}
\end{equation}
Here $g$ and $g'$ are the $SU(2)_L$ and $U(1)_Y$
gauge coupling constants, respectively; $G_{3f}$ and
$G_{2f}$ are the Yukawa
couplings between the Higgs triplet and doublet, and the
matter fermions; $N_C$ is 3 for quarks and 1 for
leptons.
$V_T = V_0 + \Delta V$ is bounded from below if
$\lambda_3 > 0, \; \lambda_2 > 0$
and $a > - \sqrt{\lambda_2 \lambda_3}.$ We make the
simplifying but unnecessary assumption that the full symmetry of the
theory (here $SU_L(2) \times U_Y(1) \times Z_2$) is restored at
high temperature.  This requires
$
\beta_3 > 0 \; \; {\rm and} \; \; \beta_2 > 0.
$
Spontaneous symmetry breaking is signaled when
$T^2$ drops below either $\mu_2^2/ \beta_2$ or $ \mu_3^2/ \beta_3$.

Now we would like to find the minima of the potential for different
temperatures and, in that way, find the symmetry breaking pattern.
There are four different phases that can occur:
Phase 1:  $<| \Phi | > =  0\ , \ \ \  <|\vec \chi | > = 0$;
Phase 2:  $<|\Phi |  > =  0 \ , \ \ \ <|\vec \chi | > \ne 0$;
Phase 3:  $<|\Phi  | > \ne  0\ , \ \ \ <|\vec \chi |> \ne 0$;
Phase 4:  $<|\Phi |> \ne  0\ ,  \ \ \ <|\vec \chi |> = 0$.
As shown in a similar model with a complex singlet instead
of the triplet \cite{kephart}, there is a range of parameters in
which the different phases occur in succession as we go from very
high to low temperatures. We have explicitly checked the existence
of such parameters for the model presented here. For example, one
such set is:
$$
g=0.65\ ,\ \  g'=0.34\ , \ \ \mu_2^2 = 0.10\ ,\ \  \lambda_2 = 0.07\ ,
\ \ \mu_3^2 = 0.22\ ,\ \  \lambda_3 = 0.48\ ,
$$
$$
a = 0.32\ {\rm and}\
b = 0\
$$
where $\mu_2$ and $\mu_3$ are measured in units of 246 GeV - the
electroweak scale.
In this case we find that Phases 1, 2, 3 and 4 follow in succession
as we cool from very high to low temperatures. In Fig. 1 we schematically
show the vevs of $\Phi$ and $\vec \chi$ as functions of temperature.
This then yields the following symmetry breaking pattern:
\begin{eqnarray}
SU(2)_L \times U(1)_Y \times Z_2 &\rightarrow &
U(1)' \times U(1)_Y \times Z_2 \nonumber \\
&\rightarrow & U(1)_Q \nonumber \\
&\rightarrow & U(1)_Q \times Z_2 \nonumber \\
\end{eqnarray}
where in the last stage the vev of $\vec \chi$ vanishes
and this restores the $Z_2$ symmetry. The group $U(1)'$ is a
subgroup of $SU(2)_L$. The generator $Q$ of the electromagnetic
group $U(1)_Q$ is a linear combination of $T'$ - the generator of
$U(1)'$ and $Y$ - the generator of $U(1)_Y$:
$$
Q = {{T' + Y} \over 2} \ .
$$
The first stage of symmetry breaking now produces topological
magnetic monopoles
in the way they would be produced in the Georgi-Glashow model
\cite{georgi}. The flux emanating from these monopoles is of the
$U(1)'$ magnetic field. At the next symmetry breaking, a fraction
of this flux becomes massive and gets squeezed into flux tubes - these
are the $Z-$strings attached to the monopoles - while the remaining
flux is massless electromagnetic flux - this is
the magnetic flux emanating from electroweak monopoles.
During this second symmetry breaking a distribution of closed
loops of $Z-$string will also be produced. Note that these strings can
break by the formation of monopole pairs but this process is exponentially
suppressed since the monopoles are assumed to be very heavy compared
to the electroweak scale. The breaking of the $Z_2$ symmetry group in the
second stage produces domain walls but these will dissolve in the last
stage when the $Z_2$ is restored.

This scheme can readily be put in a Grand Unification model because
triplets occur quite naturally. For example, in minimal $SU(5)$
grand unification, the breaking to $SU(3) \times SU(2)_L \times U(1)_Y$
is accomplished with a Higgs ${\bf 24}$; twelve of these are eaten as the
$X$ and $Y$ bosons become massive; of the remaining twelve Higgses
three are in an $SU(2)_L$ triplet that can be identified with
the $\vec \chi$ field.

We can now discuss the structure of the defects in this model and
the possible cosmological consequences. The monopoles that are
produced in the first phase transition are hedgehogs purely in the
$\vec \chi$ field and have a mass:
\begin{equation}
M_m \sim 4\pi {{m_W} \over {g^2}}
\label{mmass1}
\end{equation}
where,
\begin{equation}
m_W = \sqrt{2} g <|\vec \chi | >
\label{wmass1}
\end{equation}
is the $W$ mass in Phase 2. When the second symmetry breaking occurs,
the monopoles get connected by strings. At this stage, the $W$ boson
mass gets a contribution from the vev of $\Phi$ and is:
\begin{equation}
m_W = g \sqrt{
        {<|{\Phi}|>^2 + 4 <|{\vec \chi}|> ^2} \over 2
           }
\label{wmass2}
\end{equation}
Therefore the mass of the monopole also changes as given by
eqn. (\ref{mmass1}). The strings are topological since they
arise from the breaking of $U(1)' \times U(1)$ to $U(1)_Q$.
Apart from instanton processes resulting in the nucleation of
monopole antimonopole pairs along the string \cite{jpav}, the
strings are stable. The second symmetry breaking also yields
domain walls. We expect that
the angle between the two vectors $\vec \chi$ and
$\Phi^{\dag} {\vec \tau} \Phi$ varies as we traverse one of these walls
though, depending on the parameters of the model, one of the two
fields $\vec \chi$ or $\Phi$ could vanish in the interior of the
wall. Such walls are known to exist in condensed matter systems such
as $^3$He. Here we shall not consider the domain walls in greater
detail since they will go away
once the $Z_2$ symmetry is restored in the last stage.

In the last transition occuring at $T_3$, the vev of $\vec \chi$ vanishes.
(This feature of the model means that it is unconstrained by the
experimental limits on the $\rho$ parameter.) At temperatures below
$T_3$, the strings
are nearly identical to the $Z-$strings discussed in the literature. There
is however one difference which could be crucial - the strings can
have a bosonic condensate \cite{ew} of $\vec \chi$ on them depending
on the parameters of the model. To see this note that $\Phi$ vanishes
at the center of the string and so it is energetically favorable for
$\vec \chi$ to be non-vanishing in this region.
This means that the region in the vicinity of the strings
resembles the phase where both $\Phi$ and $\vec \chi$
have non-zero vevs - a phase where the strings are topologically
stable. Then there are two parameter dependent possibilities:
(i) the $Z$-strings are unstable at temperatures below $T_3$, and,
(ii) the $Z$-strings are stabilized at temperatures below $T_3$ by
the $\vec \chi$ condensate. Both of these possibilities may have
interesting cosmological consequences.

It is important to remember that the standard model $Z$-string is
superconducting \cite{ew} due to fermion zero modes \cite{mewp, comment3}
and can carry a maximum electric current of about $10^8$ Amps.
In our case, the fermionic part of the Lagrangian in eqn. (\ref{lag})
is unaffected by the presence
of the $\vec \chi$ field and the Dirac equations for leptons and quarks
are identical to those in the standard model. Therefore zero
modes for these particles will also exist in the
extended model $Z-$string and these strings too will be able to carry up to
$10^8$ Amps of electric current. Due to the presence of the quark
zero modes, linked and twisted
$Z-$strings will carry baryon number exactly as they do in the
standard model \cite{tvgf, jgtv}.

Now let us discuss the first possibility where the monopoles get
connected by strings at $T_2$ and then the strings, being unstable,
break up and decay by the nucleation of monopole pairs at a temperature
below $T_3$. This scenario might be interesting in the context
of baryogenesis from string decay.

The monopoles that form at $T_3$ are specified by their location,
velocity, charge and, most importantly for us, by a $U(1)'$ phase.
(It is well known that the phase of a single monopole is arbitrary
since we are free to perform $U(1)'$ rotations. But in a situation
where there are two or more monopoles, the {\it relative} phase of the
monopoles cannot be changed by such a global transformation.)
At $T_2$, $\Phi$ gets a vev and then the $\Phi$ field configuration
of a monopole and an antimonopole will have the following
asymptotic forms:
\begin{equation}
\Phi = \pmatrix{ cos(\theta /2) \cr sin(\theta /2) e^{i\phi }} \ ,
\ \ \
{\bar \Phi} = e^{i\gamma} \pmatrix{ sin({\bar \theta} /2) \cr
                          cos({\bar \theta} /2) e^{i{\bar \phi}} }
\label{phiphibar}
\end{equation}
where, $(\theta , \phi )$ and $({\bar \theta}, {\bar \phi})$
are spherical angular coordinates centered on the monopole and antimonopole
respectively and the phase $\gamma$ is the arbitrary $U(1)$ phase.
These monopoles will get connected by $Z-$strings below $T_2$
and, as shown in \cite{tvgf}, the monopoles plus string field configuration
carries baryon number. At the time the monopoles get connected by
strings, we expect their relative phases ($\gamma$)
to be uncorrelated and so each
connected monopole pair will carry of order ${\pm 1}$ baryon number.
Therefore the magnitude of the local baryon number
density at $T_2$ is of the order of the monopole density:
\begin{equation}
n_B (T=T_2) \sim \pm ~ n_m (T=T_2) \ .
\label{nbnm}
\end{equation}

If the strings are unstable at low temperatures, they will break up
into small segments below $T_1$ and these
small segments will ultimately release their energy into radiation.
As this decay happens relatively late, the strings are out of
thermal equilibrium and each decay leads to the generation of
baryon (or antibaryon) number. A net amount of baryon number will
be produced if the CP violation effects of the KM matrix are
included in the string decay. The strength of the CP violation in
the monopole-string system is not yet known but,
as the effect is non-perturbative, it need not be small. This subject
deserves further investigation.

A more exciting possibility is that the strings are stable even
at low temperatures due to the $\vec \chi$ condensate
\cite{mjlptv, tvrw}.
Then the strings do not break up into small segments below $T_1$.
The length distribution of strings is expected to be
exponentially suppressed since on following a particular string -
treated as a Brownian walk - there is a certain non-vanishing
probability that it will encounter a monopole and terminate. But
the length scale above which the exponential fall off sets in can
be quite large since it is governed by the monopole number density
and the string density at $T_2$. If the monopoles are quite dilute
when the strings form, we expect a nearly scale invariant distribution
of loops to be formed together with a few open string segments
\cite{comment2}.

The evolution of a superconducting string tangle has been discussed in
several contexts in \cite{scevoln} and the results are reviewed in
\cite{book}. The short string segments and loops
will collapse and decay into radiation with the production of baryons.
But, in the early universe, the longer string segments which carry
currents will be frozen in the ambient plasma provided their curvature
scale is larger than a critical
scale. Such strings will be stretched with the Hubble expansion
and will become even longer.  On curvature scales less than the critical
scale, the strings are free to oscillate. Such sections of strings will cut
across the random magnetic field that is also expected to be produced at
$T_2$ \cite{tvmag} and will result in an oscillating induced current on the
string \cite{aryal}.

By the epoch of structure formation, most of the strings would have
decayed but we can expect a few stray, frozen-in loops to survive.
These stray loops
would then get trapped in protogalaxies and in protostars. The
ensuing galactic and stellar turbulence would generate more string in a way
reminiscent of the generation of vortex lines in superfluids. (There
too a few stray lines result in a fully developed tangle in the presence
of flow \cite{schwarz}.)

The energy emission from superconducting strings (formed at the
electroweak scale) in the galaxy was
discussed in \cite{scevoln} and it was found that the luminosity
due to all the strings would be $\sim 10^{40}$ ergs/s. This is
significant but a characteristic signature of radiation from
strings would further assist in observations. Such a characteristic seems
possible in the case of electroweak strings due to the role that
$Z-$strings can play in baryon number violating processes. Consider,
for example, the scenario where a tangle of electroweak strings is
present in a galaxy.
(Similar phenomena would also occur in stars.)
On curvature scales smaller than
$R_* \sim 10^{16}$ cm, the strings are not frozen into the galactic
plasma. Loops on this scale will cut across the galactic magnetic field
and collapse on a time scale $\tau \sim R_* /v_s \sim 10^{11}$ sec where
$v_s \sim 10^5$ cm/s is the terminal velocity of the strings
moving through the galactic plasma \cite{scevoln}.
During this process an electric field is induced along the string and
hence $\int d^3 x {\vec E}_A \cdot {\vec B}_Z \ne 0$. The baryon
number anomaly equation then gives us the rate of baryon number
production during loop collapse \cite{barriola}. Order of magnitude
estimates tell us that such a collapsing loop releases about
$10^{-2} F_A F_Z \sim 10^{25}$ baryons where $F_A$ is the magnetic
flux that the string cuts across and $F_Z$ is the $Z-$flux in the
string. While the baryon
number production due to a single loop is quite a slow process, the
baryon number produced by all the loops in the galaxy can be
significant - about $10^{44}$ baryons in $10^{11}$ secs.
Roughly half the baryons will be in the form of antimatter and
since the strings are neutral, would be in the form of antineutrons
which then decay into antiprotons, electrons and neutrinos.
This fact may eventually be useful in the search for electroweak strings
in our galaxy \cite{lmktv}.

\section*{Acknowledgements}

We would like to thank Alan Bray, Lawrence Krauss, Glenn Starkman
and Rainer Trebin for useful comments and are grateful
to the Isaac Newton Institute for their hospitality and
for providing a unique environment where
this work was initiated. TV is grateful to the Rosenbaum Foundation
for support during his stay at the Isaac Newton Institute.
The work of TWK was supported by the U. S. Department of Energy grant
DE-FG05-85ER40226 and the Vanderbilt University
College of Arts and Sciences.

\section*{Figure captions}

\begin{itemize}

\item{\bf Fig. 1} A schematic diagram of the vacuum expectation values
of $\Phi$ and $\vec \chi$ as functions of temperature. When the temperature
falls to $T_1$, $\vec \chi$ gets a vev. At $T_2$, $\Phi$ starts acquiring
a vev and the vev of $\vec \chi$ starts turning off and vanishes at
$T_3$.

\end{itemize}


\begin{thebibliography}{999}
\bibitem{comment0} If $Z_2$ is absent, the cubic term in the
potential is allowed and is constrained to be small by the experimental
results on the $\rho$ parameter.
\bibitem{comment3} If the $\vec \chi$ field forms a bosonic condensate
on the string, there will also be a contribution to the electric current
from the bosonic charge carriers on the string. This additional current
will contribute to the energy and quantum numbers on linked loops of
string as discussed in \cite{jgtv}.
\bibitem{comment2} There
is also the possibility that, at a critical dilution of monopoles,
the strings will percolate and a few very large strings
will form.
\end{thebibliography}
\end{document}